\documentstyle[12pt]{article}
\begin{document}
\def\slash#1{#1 \hskip -0.5em / }
\global\arraycolsep=2pt 
\topmargin 0cm
\textheight 22cm
\input{epsf}
\thispagestyle{empty}
\begin{titlepage}

\begin{flushright}
CERN-TH.7308/94\\
Technion-PH-94/9  \\
hep-ph/94mmnnn
\end{flushright}

\vspace{0.3cm}

\begin{center}
{\Large\bf On the Determination of $|V_{ub}|$ \\
           from Inclusive
           Semileptonic Decay Spectra}
\vspace{1cm} \\
\end{center}

\vspace{0.8cm}

\begin{center}
Boris Blok     \\
{\sl Physics Department, Technion, Haifa, Israel} \vspace{3mm} \\
Thomas Mannel  \\
{\sl Theory Division, CERN, CH-1211 Geneva 23, Switzerland}
\end{center}

\vspace{0.8cm}

\begin{abstract}
\noindent
We propose a model independent method to determine $|V_{ub}|$ from the
energy spectrum of the charged lepton in inclusive semileptonic
$B$ decays. The method includes perturbative QCD corrections as well
as nonperturbative ones.
\end{abstract}
\vfill
\noindent
CERN-TH.7308/94\\
June 1994
\end{titlepage}
\section{Introduction}
Recently the inclusive decay spectra of heavy hadrons have
attracted renewed attention. It has been shown that by a
combination of Heavy Quark Effective Theory and the Operator
Product Expansion one may perform a systematic $1/m_Q$ expansion
of the charged lepton spectrum in inclusive heavy hadron decays
\cite{CDG} - \cite{M}.
However, it turns out that an adequate description of the endpoint
region, where the charged lepton energy
 $E_\ell \sim E_{max}$, requires a partial resummation of the
Operator Product Expansion, yielding a result analogous to the
leading twist terms in deep inelastic scattering. In particular,
it involves the analogue of the parton distribution function, which
in the present case describes the endpoint region of the lepton
spectrum. While in most of the phase space the nonperturbative
corrections may be described in terms of a few parameters, the endpoint
region needs the input of a nonperturbative function
\cite{Neubert}-\cite{MN}.
These recent ideas constitute a substantial conceptual progress towards
a model independent description of inclusive decay spectra, including
the endpoint region.

However, aside from the nonperturbative effects
one also has to include QCD radiative corrections before one may
confront the theory with data. These corrections have been calculated
\cite{Ali}-\cite{Jesabek}
and are known to be small for $b \to c$ transitions over the whole phase
space. This is also true for the $b \to u$ case, as long as one is
not too close to the endpoint $E_\ell \sim E_{max}$. Neglecting the mass
of the $u$ quark one finds Sudakov-like double logarithms of the form
$\ln^2 (E_{max}-E_\ell)$ as well as single logarithms
$\ln (E_{max}-E_\ell)$, indicating a breakdown of perturbation theory
close to the endpoint.

In the present context the QCD radiative corrections have been studied
by Falk et al.\cite{Falk}
and Bigi et al.\cite{Bigi1}. Falk et al. arrive at the conclusion
that the endpoint region of the $b\to u$ decays is strongly affected
by QCD radiative corrections and in principle a resummation to all orders
becomes necessary.
The conclusion
of \cite{Falk} is that the extraction of
the nonperturbative effects becomes practically
impossible due to large and mainly uncontrollable perturbative effects.

In \cite{Bigi1} the Sudakov logarithms are exponentiated
leading to a strong modification of the endpoint region.
The authors of \cite{Bigi1} suggest that
uncontrollable radiative corrections discussed in \cite{Falk}
may be eliminated by the appropriate
choice of the scale $\mu$ entering the strong coupling.
They use $\mu \sim \Lambda m_b$,
where $\Lambda $ is the characteristic momentum of light degrees of
freedom in the heavy meson.
However, this leads to a strong modification of the shape of the
spectrum near the endpoint and the reliability
of the calculation becomes questionable.

Unfortunately, it is only the endpoint region of charmless
$B$ decays, which is not buried in the huge background
of the charmed decays. The kinematic endpoint of the charmed $B$ decays
is $(m_B^2 - m_D^2)/(2m_B) \sim $ 2.3 GeV, while for the charmless
semileptonic decay it is $(m_B^2 - m_\pi^2)/(2m_B) \sim$ 2.6 GeV. Hence
there is only a window of at maximum 300 MeV, which may be attributed
solely to the transition $b \to u \ell \nu$. The size of this window is
of the order of the mass difference of the heavy meson and the heavy
quark, $\bar\Lambda = m_B - m_b$, and thus it is not dominated by
a few resonances. This would be the case in a smaller region of the
size $\bar\Lambda (\bar\Lambda / m_b) $, which is of the order of
tens of MeV, and in which methods mentioned above would fail,
since they are based on parton hadron duality.

The present paper is an attempt to define quantities which are on
one hand experimentally accessible, and on the other hand allow to
disentangle perturbative and nonperturbative effects. Our suggestion
is to calculate appropriate moments ${\cal M}_n$  of the measured
spectrum, which are defined in such a way that they allow an
expansion in both $\alpha_s (m_b)$ and $\bar\Lambda / m_b$, at least
for some range of indices $n$.

Similar ideas have been put forward previously in \cite{Neubert} and
\cite{Bigi1}. However, in these papers there has been no detailed
numerical study of such moments, including both radiative and
nonperturbative effects.

In the next section we shall reconsider the perturbative and
nonperturbative contributions to the decay spectrum of inclusive
semileptonic charmless $B$ decays and give the definition of the
moments. In section 3 we perform a numerical study
from which we obtain constraints on the range of the index $n$
of the moments. Finally we give our conclusions and comment on
the extraction of $V_{ub}$.

\section{Perturbative and Nonperturbative Contributions to the
         Spectrum}
It has been shown in \cite{CDG} -\cite{M}
that one may obtain a systematic $1/m_b$ expansion
for inclusive decay rates and the corresponding spectra.
The leading term in this expansion is
the free quark decay, and the first nonvanishing corrections
appear formally at the order $1/m_b^2$. These are genuinely
nonperturbative corrections, which are given in terms of two
matrix elements
\begin{eqnarray}
&& \langle B(v) | \bar{h}_v (iD)^2 h_v | B(v) \rangle = 2m_B \lambda_1
\\
&& \langle B(v) | \bar{h}_v (-i)
   \sigma_{\mu \nu} (iD^\mu) (iD^\nu) h_v | B(v) \rangle =
   6m_B \lambda_2 .
\end{eqnarray}
The parameter $\lambda_2$ is given in terms of the $0^- - 1^-$ mass
splitting $ \lambda_2 = (m_{B^*}^2 - m_B^2)/4
\sim 0.12$ GeV$^2$. The matrix element $\lambda_1$ is not as easily
accessible, in particular it has not yet been determined experimentally.
The theoretical estimates vary over a broad range \cite{N,B} of about
$\lambda_1\sim -(0.3 -0.6)$ GeV$^2$; thus
we shall consider below a typical ``small'' value $\lambda_1 = -0.3$
GeV${}^2$ and a ``large'' one, $\lambda_1 = -0.6$ GeV${}^2$.

The result for the spectrum of the inclusive decay $B \to X_u
\ell \nu$, neglecting the mass of the $u$ quark, is given by
\begin{eqnarray} \label{npc}
\frac{1}{\Gamma_b} \frac{d\Gamma}{dy} &=& \left[ 2y^2 (3-2y)
        + \frac{10y^3}{3} \frac{\lambda_1}{m_b^2}
        + 2y^2(6+5y) \frac{\lambda_2}{m_b^2} \right] \Theta(1-y)
\\ \nonumber
 && \quad - \frac{\lambda_1 + 33 \lambda_2}{3m_b^2} \delta (1-y)
  -  \frac{\lambda_1}{3m_b^2} \delta ' (1-y))  , 
\end{eqnarray}
where
\begin{equation}
y = \frac{2 E_\ell}{m_b} \qquad
\Gamma_b = \frac{G_F^2 |V_{ub}|^2}{192 \pi^3} m_b^5  .
\end{equation}
The $\delta$ function singularities at the parton model endpoint
$y = 1$ indicate a breakdown of the operator product expansion. In
fact, it has been pointed out in \cite{BBSUV,BSUV,BKSV} that for the
decay spectra the expansion parameter is in fact $\bar\Lambda / [m_b (1-y)]$,
which becomes large close to the endpoint. Still these terms
have an interpretation as being the first few terms of a
moment expansion of a nonperturbative function describing
the behavior of the spectrum close to the endpoint \cite{Neubert}-\cite{MN}:
\begin{equation}\label{eq:contraction}
\frac{d\Gamma}{dy}  =  \Gamma_b \left[2 \Theta (1-y)
                     + \sum_{n=1}^\infty \frac{a_n}{m_Q^n} \frac{1}{n!}
                        \delta ^{(n-1)} (1-y) \right]  
 =  2 \Gamma_b \int^1_{(y-1)m_b/\bar\Lambda}F(x)dx  , 
\end{equation}
where $\delta^{(k)}$ denotes the $k$th derivative of the $\delta$
function. Note that $a_1$ vanishes and $a_2$ may be read off from
(\ref{npc}) to be $a_2 = -\lambda_1 / 3$.

The function $F(x)$ may be written formally as
\begin{equation} \label{npfunc}
F(x) = \frac{1}{2 m_B}
\langle B(v) | \bar{h}_v
\delta \left( x- \frac{iD_+}{\bar\Lambda} \right) h_v | B(v) \rangle  ,
\end{equation}
where $D_+$ is the positive light cone component of the covariant derivative.
This function has support only in a region of $x \sim {\cal O} (1)$.
In particular, the function $F$ leads to some ``smearing''
of the endpoint region, with the effect that the endpoint of the
spectrum is shifted from the parton model endpoint $m_b / 2$ to the
physical endpoint $m_B / 2$, where $m_B$ is the $B$-meson mass.
Below we shall use some model input for $F$ to estimate the size of the
higher order corrections close to the endpoint.

Aside from these nonperturbative corrections there are  also
perturbative ones, which have been calculated some time ago.
The order $\alpha_s$ corrections are known for all values of
the lepton energy \cite{Ali}-\cite{Jesabek};
however, since we are interested only in the
behavior close to the endpoint, we shall consider here only the
contributions relevant in this region. It turns out, that the
order $\alpha_s$ corrections exhibit doubly and singly logarithmic
divergencies at the endpoint. Up to terms vanishing at the endpoint,
the result reads \cite{Jesabek}
\begin{equation} \label{rc}
\frac{d\Gamma}{dy} = \frac{d\Gamma^{(0)}}{dy}
\left[ 1 - \frac{2 \alpha_s}{3 \pi} \left(
\ln ^2 (1-y) + \frac{31}{6} \ln (1-y) + \frac{5}{4} + \pi^2 \right) \right],
\end{equation}
where at the endpoint
\begin{equation} \label{endpoint}
\frac{1}{\Gamma_b} \frac{d\Gamma^{(0)}}{dy} = 2y^2 (3-2y) \Theta(1-y)
\to 2 \Theta(1-y) .
\end{equation}

 The scale $\mu$ of the strong coupling is not yet fixed in this
expression; any scale choice
 will formally only affect subleading terms. However,
 from physical considerations one may be lead to chose
$\mu \sim m_b^2 (1-y)$ as it was done in \cite{Bigi1}\footnote{
    Actually, the scale chosen in \cite{Bigi1} is
    $\mu^2 \sim \bar\Lambda m_b$, but this is equivalent to
    $\mu \sim m_b^2 (1-y)$, since in the endpoint region
    $1-y\sim \bar\Lambda/m_b$.},
leading to an
even more singular behavior as $y \to 1$. In any case, a resummation
of the singular terms becomes mandatory in order to describe the spectrum
close to the endpoint.

The common folklore is that the doubly logarithmic terms in
(\ref{rc}) exponentiate, and at the level of the Sudakov logarithms
the radiative corrections become
\begin{equation} \label{rc2}
\frac{d\Gamma}{dy} = \frac{d\Gamma^{(0)}}{dy}
\exp\left[-\frac{2 \alpha_s }{3 \pi} \ln ^2 (1-y)\right]  .
\end{equation}
Again the problem arises of the scale of $\alpha_s$  and we shall
compare below two choices. Using the one loop expression for
$\alpha_s$
\begin{equation}
\alpha_s (\mu ) = \frac{12 \pi}{(33-2n_f) \ln(\mu^2 / \Lambda_{QCD}^2)} , 
\end{equation}
we will compare the results obtained for $\mu = m_b$ and
$\mu \sim m_b^2 (1-y)$, the latter expression leading to
\begin{equation} \label{rc3}
\frac{d\Gamma}{dy} = \frac{d\Gamma^{(0)}}{dy}
\exp\left[-\frac{8}{25} \frac{\ln ^2 (1-y)}
                        {\ln[(m_b^2 / \Lambda_{QCD}^2) (1-y)]} \right] .
\end{equation}
In the endpoint region  one obtains qualitatively a behavior
of the form \cite{Bigi1}
\begin{equation}
\frac{d\Gamma}{dy} \sim \frac{d\Gamma^{(0)}}{dy} (1-y)^{(\epsilon_0 -1)} ,
\label{Bigi1}
\end{equation}
where
$$
\epsilon_0 -1= -\frac{8}{25} \frac{\ln (1-y)}
                        {\ln[(m_b^2 / \Lambda_{QCD}^2) (1-y)]}
$$
behaves approximately like a constant in the region of interest
\cite{Bigi1}. However, after such a resummation to all orders
of perturbation theory, it is not obvious, how to disentangle perturbative
and nonperturbative contributions any more. This becomes clear, if one
really treats $\epsilon_0$ as a constant, in which case one may rewrite
(\ref{Bigi1})
\begin{equation}
\frac{d\Gamma}{dy} \sim \frac{d\Gamma^{(0)}}{dy}
\mbox{ const. }
 \exp\left\{\frac{12\pi (\epsilon_0 -1)}{25}
                   \frac{1}{\alpha_s [(m_b^2 / \Lambda_{QCD}^2) (1-y)]}
                  \right\} , 
\label{Bigi1a}
\end{equation}
which in this form looks like a nonperturbative contribution.

The above discussion shows that the endpoint region is in fact
difficult to describe. On one hand there are large nonperturbative
contributions, on the other hand
there are perturbative contributions, becoming large in the
endpoint region. These remarks, however, apply only to the spectrum itself.
It has already been shown in \cite{Neubert,Bigi1} that it may be
helpful to consider appropriately defined moments of the spectrum; e.g.\
the moments of the spectrum taken with respect to the parton model
endpoint are related to the
coefficients of the most singular $\delta$ function like singularities
at the endpoint, in each order in the $1/m_b$ expansion.

Here we propose to consider a  slightly different set of moments,
namely
\begin{equation}
{\cal  M}_n = \int\limits_0^{1+\bar\Lambda / m_b}  dy  \,\,
 y^n \frac{d\Gamma}{dy} , 
\end{equation}
which may be rewritten in terms of the moments as defined in
\cite{Neubert}. As we shall discuss in detail below,
these moments have, for some range of $n$,
a simultaneous expansion in $\alpha_s (m_b)$
and $\bar\Lambda / m_b$.

This range of $n$ is specified by two requirements. Experimental
information will be available only close in the endpoint region,
and thus $n$ has to be large enough to be sensitive to this region.
On the other hand, the larger $n$ is, the stronger the sensitivity to
the details of the endpoint region is. In other words, only for
not too large $n$ one may perform the simultaneous perturbative
and nonperturbative expansion; the problems present in the
spectrum will reappear in the behavior of the moments
${\cal M}_n$ for large $n$.

In fact, from order of magnitude considerations one would expect
that there is no such range in $n$, where $n$ is large enough
to be sensitive to the endpoint, and in which nonperturbative and
radiative effects still remain under control. However, it turns
out from the more detailed study presented below that there might be
such a a window in $n$.

In the next section we shall perform a numerical study
for the radiative corrections and the nonperturbative contributions,
in order to find a range in $n$ where one may reliably calculate the
moments ${\cal M}_n$.

\section{Numerical Discussion}
\begin{figure}[t]
   \vspace{0.5cm}
   \epsfysize=9cm
   \centerline{\epsffile{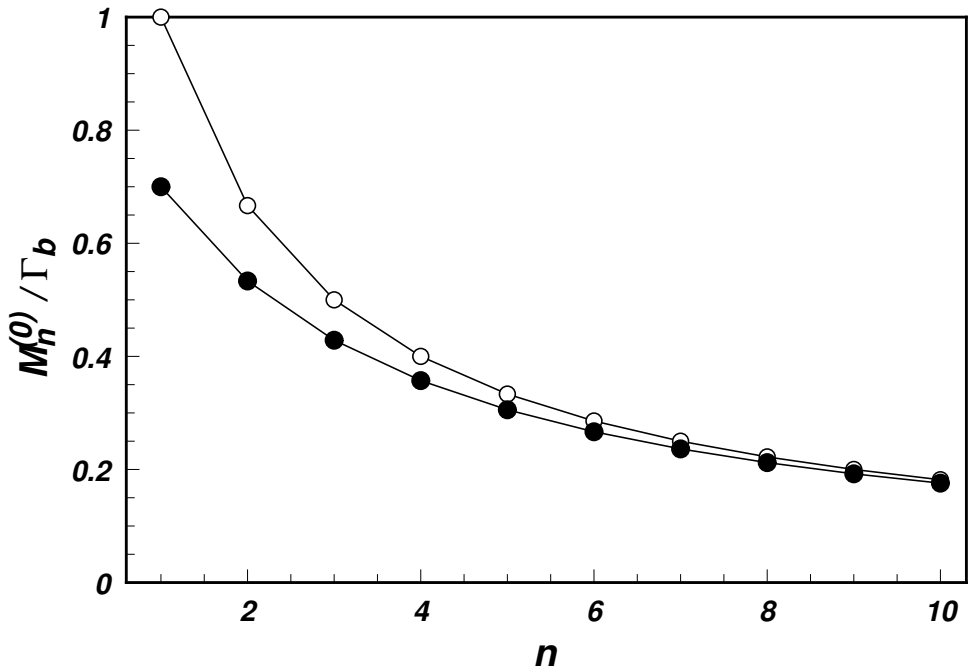}}
   \centerline{\parbox{11cm}{\caption{\label{fig1}
Comparison of the parton model moments for the full $y$ dependence
(full dots) versus the endpoint approximation (open circles).   }}}
\end{figure}

We shall first consider the moments of the differential
distribution in the naive parton model, without any
corrections. The spectrum is given by the parton model expression
and one obtains for the moments to the zeroth order
\begin{equation}
{\cal  M}_n^{(0)} = \Gamma_b \frac{2(n+6)}{(n+3)(n+4)}
\end{equation}
However, we are mainly interested in the endpoint region, where the
parton model spectrum may be approximated by (\ref{endpoint}), and the
moments obtained in this approximation are
\begin{equation} \label{m0approx}
\tilde{{\cal M}}_n^{(0)} = \Gamma_b \frac{2}{n+1}
\end{equation}
In Fig.\ref{fig1} we plot the moments obtained from the full
parton model rate versus the approximation (\ref{m0approx}). The
comparison between the two sets gives some impression, at what
$n$ one is sensitive only to the endpoint region; from the figure
one reads off that already at $n \sim 4$ one mainly obtains information
on the endpoint.

On the other hand, data will be available in the near future only for
lepton energies above 2.3 GeV. Thus we shall in the following also
consider moments, in which the integration over $n$ is restricted to
a range $y_0 < y < 1+\bar\Lambda / m_b$
\begin{equation} \label{mcut}
{\cal  M}_n (y_0) = \int\limits_{y_0}^{1+\bar\Lambda / m_b}  dy  \,\,
 y^n \frac{d\Gamma}{dy}  ,
\end{equation}
where realistic values for $y_0$
will be in the region of $y_0 = 0.9$. Introducing such a lower cut will
change (\ref{m0approx}) to
\begin{equation} \label{m0a1}
\tilde{{\cal M}}_n^{(0)} = \Gamma_b  \frac{2}{n+1}
                           \left(1-y_0^{n+1}\right) ,
\end{equation}
from which we estimate that for $y_0 = 0.9$ the sixth moment will
get about $52\%$, the tenth moment already about $70\%$ contribution
from the region $y_0 < y < 1$, at least in the naive parton model.
On the other hand, an upper limit of $n$ is given by
the experimental resolution. If data on $b \to u$ semileptonic transitions
is available only in the small window between 2.3 GeV and 2.6 GeV one
may not expect enough data to extract moments higher than about ten.
In what follows we shall thus concentrate on moments less than ten.

Next we turn to the nonperturbative corrections.  They have been
been given in (\ref{npc})
to order $1/m_Q^2$. Taking the moments of (\ref{npc}) one obtains
in the endpoint approximation (\ref{endpoint})
\begin{equation} \label{npfull}
{\cal M}_n = \tilde{{\cal M}}_n^{(0)} + \Gamma_b
\left[ \frac{\lambda_1}{3m_b^2} \left(\frac{10}{n+1}- 1 - n \right)
       + \frac{\lambda_2}{m_b^2} \left(\frac{22}{n+1} -11 \right)
\right] .
\end{equation}
The dependence on $n$ of the various terms in ${\cal M}_n$
reflects, how singular
the contribution is at the endpoint of the lepton spectrum. The most
singular term is the one with the derivative of the $\delta$ function; this
leads to a linear dependence in $n$. The terms behaving like a
$\delta$ function yield constant terms in the moments, while
terms with step functions will decrease as $1/n$ if $n$ becomes large.

In a similar way as for the parton model we may estimate the effects
of a lower cut off $y_0$ according to (\ref{mcut}). For the
terms involving step function the
result is qualitatively the same as for the parton model, while for the
contributions with $\delta$ function like singularities there is no
dependence on the lower cut, since these are concentrated at $y=1$.

Higher moments will become more and more sensitive to what happens
in the endpoint region. Thus (\ref{npfull}) is not valid for $n$ becoming
too large. In order to estimate the value of $n$, where the expansion
breaks down, we have to consider an even more singular contribution. This
is contained in the next order of the $1/m_Q$ expansion, and has the form
\begin{equation}
\frac{d\Gamma}{dy} \sim \frac{a_3}{m_b^3} \delta '' (1-y) ,
\end{equation}
where the parameter $a_3$ is related to the matrix element
$$
\langle B(v) | \bar{h}_v (iD_\mu) (ivD) (iD^\mu) h_v | B(v) \rangle .
$$
Taking the moment of this gives a contribution quadratic in $n$
\begin{equation}
M_n \sim  \Gamma_b n (n-1) \frac{a_3}{m_b^3} . 
\end{equation}
Neglecting the higher orders in the $1/m_Q$ expansion of the moments
is only justified, if
$$
n (n-1) \frac{a_3}{m_b^3} < n \frac{\lambda_1}{m_b^2} .
$$
Estimating $a_3 $ to be $(|\lambda_1|)^{3/2} \sim (500$ MeV$)^3$
one obtains
\begin{equation}
n < \frac{m_b}{\sqrt{|\lambda_1|}} \sim 10 , 
\end{equation}
which means that only the nonperturbative contributions to the
first few moments may be calculated to lowest order in the $1/m_Q$
expansion. It is a fortunate circumstance that this upper limit in
$n$ coincides with the maximal $n$ accessible in experiment.

In order to estimate higher moments one has to perform a
resummation of the most singular terms at the endpoint to all orders
in $1/m_Q$, in other words, one has to include these effects using
the nonperturbative function $F(x)$
defined in (\ref{npfunc}). In terms of this the rate is given by
(\ref{eq:contraction}), which  describes the behavior of the
lepton spectrum close to the
endpoint, namely over a region of the order
$1-y \sim \sqrt{|\lambda_1|}/m_b$. There is, however, an even smaller
region $1-y \sim (\sqrt{|\lambda_1|}/m_b)^2$, the resonance region,
in which only a few resonances contribute to the spectrum. This region
will start to contribute significantly to the moments for
\begin{equation}
n \sim (\frac{m_b}{\sqrt{|\lambda_1|}})^2 \sim 100
\end{equation}
which means that moments with $n > 100$ will be determines
 from the contributions of only very few light resonances.

The function $F$ is genuinely nonperturbative and it has been considered
using several models. One frequently used model is the one of
Altarelli et al. \cite{ACCMM}, which has
been rewritten in terms of the nonperturbative function $F(x)$
\cite{Bigi2} given by
\begin{equation} \label{ACCMM}
F(x)=\frac{\bar \Lambda}{\sqrt{\pi} \, p_F}
 \exp\left(-\frac{1}{4} \left(\frac{\bar \Lambda\rho}{p_F(1-x)}
         - \frac{p_F}{\bar \Lambda}(1-x) \right)^2 \right) .
\label{basic}
\end{equation}
Here  $\bar \Lambda = m_B-m_b$, where $m_B$ is the mass of the
$B$-meson. The second parameter $p_F$ is the so called Fermi-momentum
which corresponds to the motion of the heavy quark inside of the
heavy meson; see \cite{Bigi2} for precise definition. Finally, $\rho$
is implicitly given in terms of the other two parameters by
\begin{equation}
\bar \Lambda =\frac{p_F\rho e^{(\rho /2)} K_1(\rho /2)}{\sqrt{\pi}} ,
\label{connection1}
\end{equation}
where $K_1$ is a modified Bessel function.

The model as given in \cite{Bigi2} has only one parameter $\rho$,
which has a physical
interpretation, namely $\rho = m_{\rm Spectator}^2 / p_F^2$. One may also
relate the model parameters to the QCD matrix element $\lambda_1$,
since appropriate  moments of $F$ are related to matrix elements
of powers of covariant derivatives between heavy meson states. In this
way one obtains the relation
\begin{equation}
1-\frac{\lambda_1}{3\bar \Lambda^2}=
\pi \frac{2+\rho}{\rho^2 K_1^2(\rho /2)} e^{-\rho} ,
\label{connection2}
\end{equation}
which may be used to determine the value of the model parameter
$\rho$ in terms of $\bar\Lambda$ and $\lambda_1$; it has
been pointed out in \cite{Bigi2} that only the
combination $\xi= -\lambda_1/(3\bar\Lambda^2)$ enters in the model.
In particular, eq.(\ref{connection2}) has no solution, if $\xi$
is larger than 0.57. For a value of $\bar\Lambda$ of 500 MeV, this means
that the model cannot accommodate a value for $-\lambda_1$ of more than
0.42 GeV${}^2$. For this value we have $\rho = 0$, and the nonperturbative
function simplifies
\begin{equation}
F(x)=\frac{2}{\pi} \exp \left(-\frac{(1-x)^2}{\pi} \right) .
\label{eq:roman}
\end{equation}
We shall use this model to estimate the effects of the most singular
terms occurring in higher orders in the $1/m_b$ expansion.

The value of $\lambda_1$ is not yet known accurately. We shall consider
below a ``small'' value $\lambda_1 = 0.3$ GeV${}^2$ as well as the largest
value possible in the above model, $\lambda_1 = 0.43$ GeV${}^2$. The first value of $\lambda_1$ corresponds to $\rho = 0.35$, while the
large one is the limiting case $\rho = 0$. In our final result we
shall also consider a ``large'' value $\lambda_1 = -0.6$GeV${}^2$.
Furthermore, we will use $m_b = 4.7$ GeV
and $\bar\Lambda\sim 500$ MeV in the following numerical analysis.

\begin{figure}[p]
   \vspace{-0.5cm}
   \epsfysize=9cm
   \centerline{\epsffile{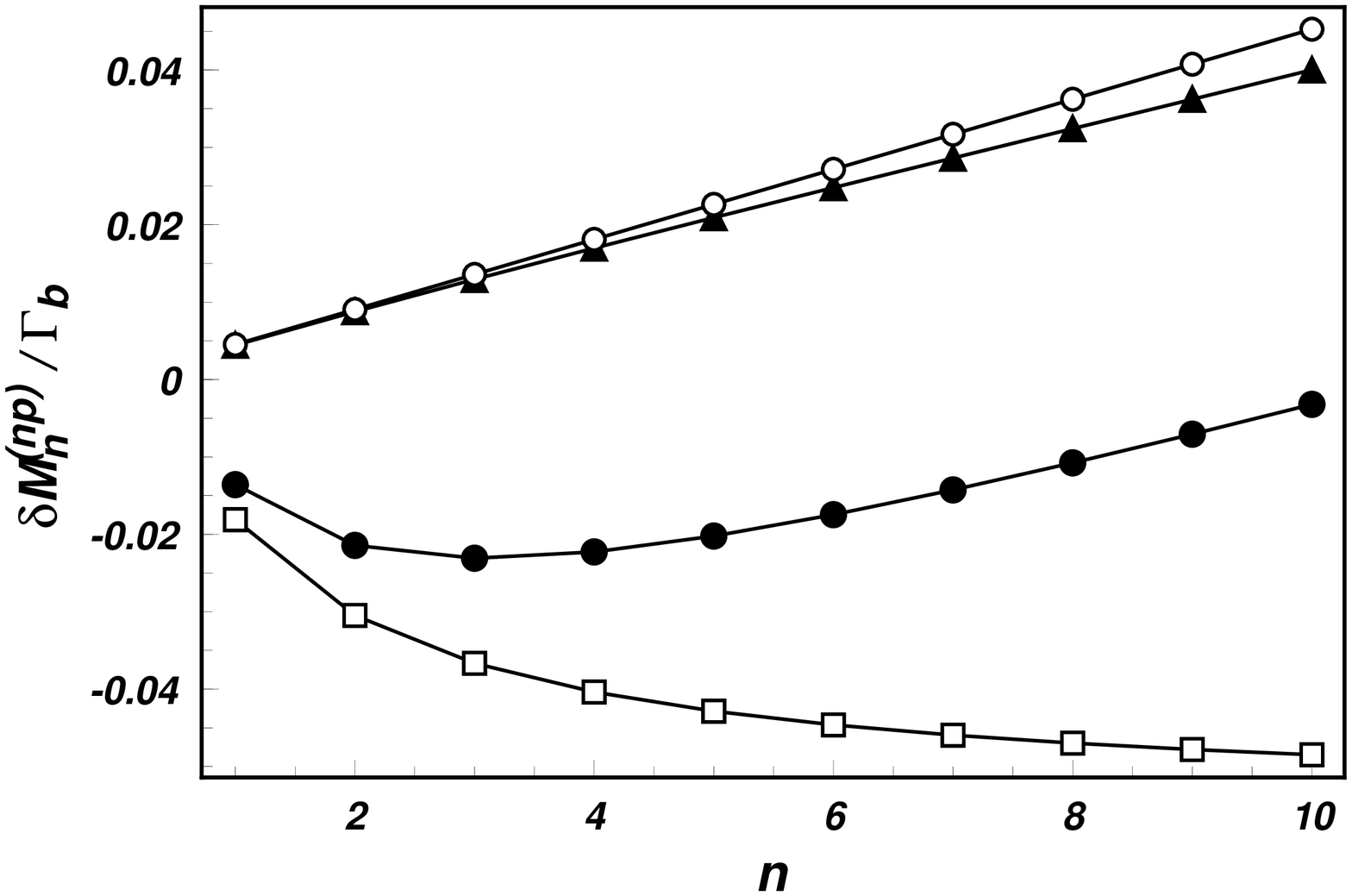}}
   \epsfysize=9cm
   \centerline{\epsffile{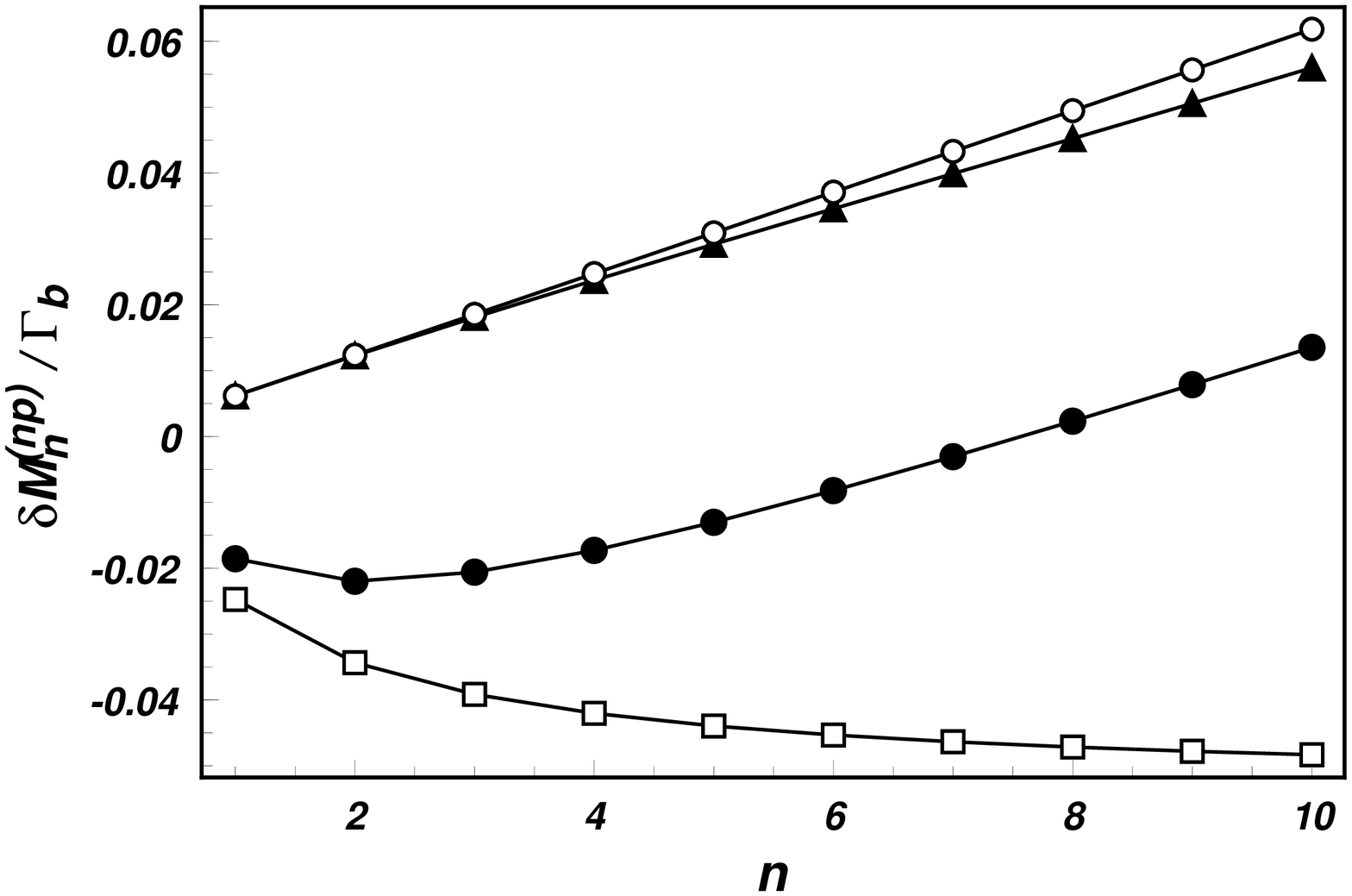}}
   \centerline{\parbox{11cm}{\caption{\label{fig2}
The nonperturbative corrections to the moments for $\lambda_1 = -0.3$
GeV${}^2$ (upper figure) and $\lambda_1 = -0.43$ GeV${}^2$ (lower figure).
The solid dots are
the corrections using the lowest nontrivial contributions
(\protect{\ref{npfull}}), the solid triangles are the corrections
from the nonperturbative function as given in (\protect{\ref{eq:roman}}),
the open circles are the contributions from the $n \lambda_1$ alone in
(\protect{\ref{npfull}}), and the open boxes are the contributions
except the $n \lambda_1$ term.
   }}}
\end{figure}

In Fig.\ref{fig2} we plot the nonperturbative corrections
$\delta {\cal M}_n^{(np)}$ to the moments, $\delta {\cal M}_n^{(np)}
= {\cal M}_n^{(np)}- \tilde {\cal M}_n^{(0)}$.
The upper plot corresponds to
$\lambda_1 = -0.3$ GeV$^2$, and the lower one to the maximal
value for $\lambda_1$ which can be accommodated in the model (\ref{ACCMM}),
$\lambda_1\sim -0.43$ GeV$^2$. The solid dots are the nonperturbative
corrections according to (\ref{npfull}), while the solid triangles are
the corrections using the model (\ref{ACCMM}) for the nonperturbative
function $F$.

We have also plotted separately  the contributions to the moments
originating from the term linear
in $n$ of (\ref{npfull}) (open circles) and the rest, i.e.\ the constant
terms and terms decreasing with $n$ (open squares).

In both cases, small and large $\lambda_1$, one finds a substantial
contribution from the chromomagnetic moment term $\lambda_2$ at least for
small moments. The term from the kinetic energy operator has a linear
dependence on $n$ (c.f. (\ref{npfull})) while the chromomagnetic term
behaves like a constant for large $n$. The linear terms of (\ref{npfull})
are well reproduced by the model for the nonperturbative function;
this is to be expected, since the nonperturbative function contains
only the most singular contribution in each order of the
$1/m_b$ expansion, i.e. the leading twist term. On the other hand, from
the fact that the linear term in $n$ from (\ref{npfull}) already
approximates the nonperturbative function quite well, one may conclude
that the expansion (\ref{npfull}) is in fact sufficient for the range
of $n$ we are considering. In other words, the leading twist contribution
modelled by the ansatz (\ref{ACCMM}) may be replaced by the term
proportional to $\lambda_1/m_b^2$ in the $1/m_b$
expansion of the moments with high accuracy.

However, the result (\ref{npfull}) differs from what the nonperturbative
function gives, mainly because of the chromomagnetic moment term and the
constant terms proportional to $\lambda_1$. These
term are less singular than the contribution of $ n \lambda_1$,
but they contribute substantially to the moments $n < 10$. On the
other hand, the leading twist terms indicate that the $1/m_b$ expansion is
satisfactory for $n < 10$ and we conclude the expansion (\ref{npfull})
is justified for the moments in the range considered.

\begin{figure}
   \vspace{0.5cm}
   \epsfysize=9cm
   \centerline{\epsffile{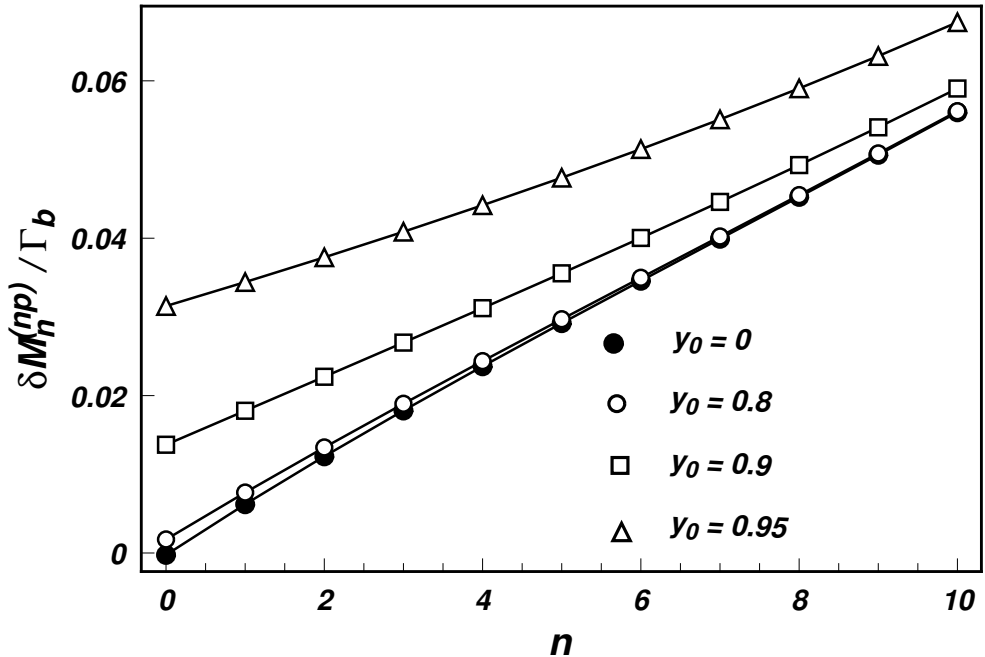}}
   \centerline{\parbox{11cm}{\caption{\label{cutrom}
The nonperturbative corrections to the moments, including a lower
cut of $y_0$ in the integration as in (\protect{\ref{mcut}}). We use
$\rho = 0$, corresponding to $\lambda_1 = -0.43$ GeV${}^2$.
 }}}
\end{figure}

Finally, one may also consider the effect of a lower cut off $y_0$
in the integration over $y$ as in (\ref{mcut}). The strongest
effects are expected for
large  $\lambda_1$, so we plot in fig.\ref{cutrom} the nonperturbative
corrections for $\rho = 0$ in the model (\ref{ACCMM}), 
for different values of the cut off $y_0$.
For the realistic case $y_0 = 0.9$ one finds only small corrections
for $4<n<10$, and the conclusion concerning the minimal $n$ obtained
from the parton model remain valid.

\begin{figure}[t]
   \vspace{0.5cm}
   \epsfysize=9cm
   \centerline{\epsffile{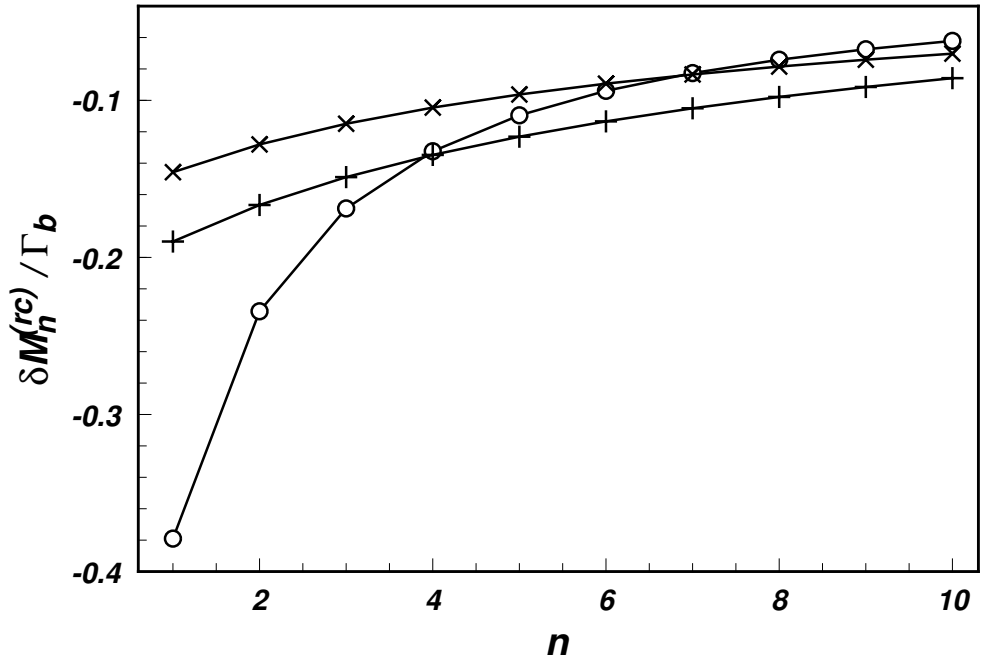}}
   \centerline{\parbox{11cm}{\caption{\label{fig3}
The radiative corrections to the moments. Open circles correspond to
(\protect{\ref{rc}}) with $\alpha_s(m_b) = 0.26$, 
the crosses correspond to (\protect{\ref{rc2}}) with $\alpha_s(m_b) = 0.26$,
and the x'es are obtained from (\protect{\ref{rc3}}).
 }}}
\end{figure}

Next we consider the radiative corrections. In Fig.\ref{fig3}
we plot these corrections to the moments
$\delta {\cal M}_n^{(rc)}$ obtained from the one
loop result (\ref{rc}),
from the exponentiation of the leading double logarithms, for both
fixed scale $\mu = m_b$ and for variable scale $\mu = m_b (1-x)$.
We use in (\ref{rc}) the value of $\alpha_s$ at $m_b$,
$\alpha_s (m_b) = 0.26$, corresponding to
$\Lambda_{QCD} = 250$ MeV.

The radiative correction contribution to the moments with $n > 4$
is rather stable, in other words, it does not depend on how one deals
with the endpoint singularities. Compared to the nonperturbative corrections
for small $\lambda_1$ they are much larger, dominating the corrections
to the moments. This smallness of the nonperturbative corrections is
due to an almost cancellation between the contributions of $\lambda_1$ and
$\lambda_2$ in the region of $n$ we are considering.

\begin{figure}[p]
   \epsfysize=9cm
   \centerline{\epsffile{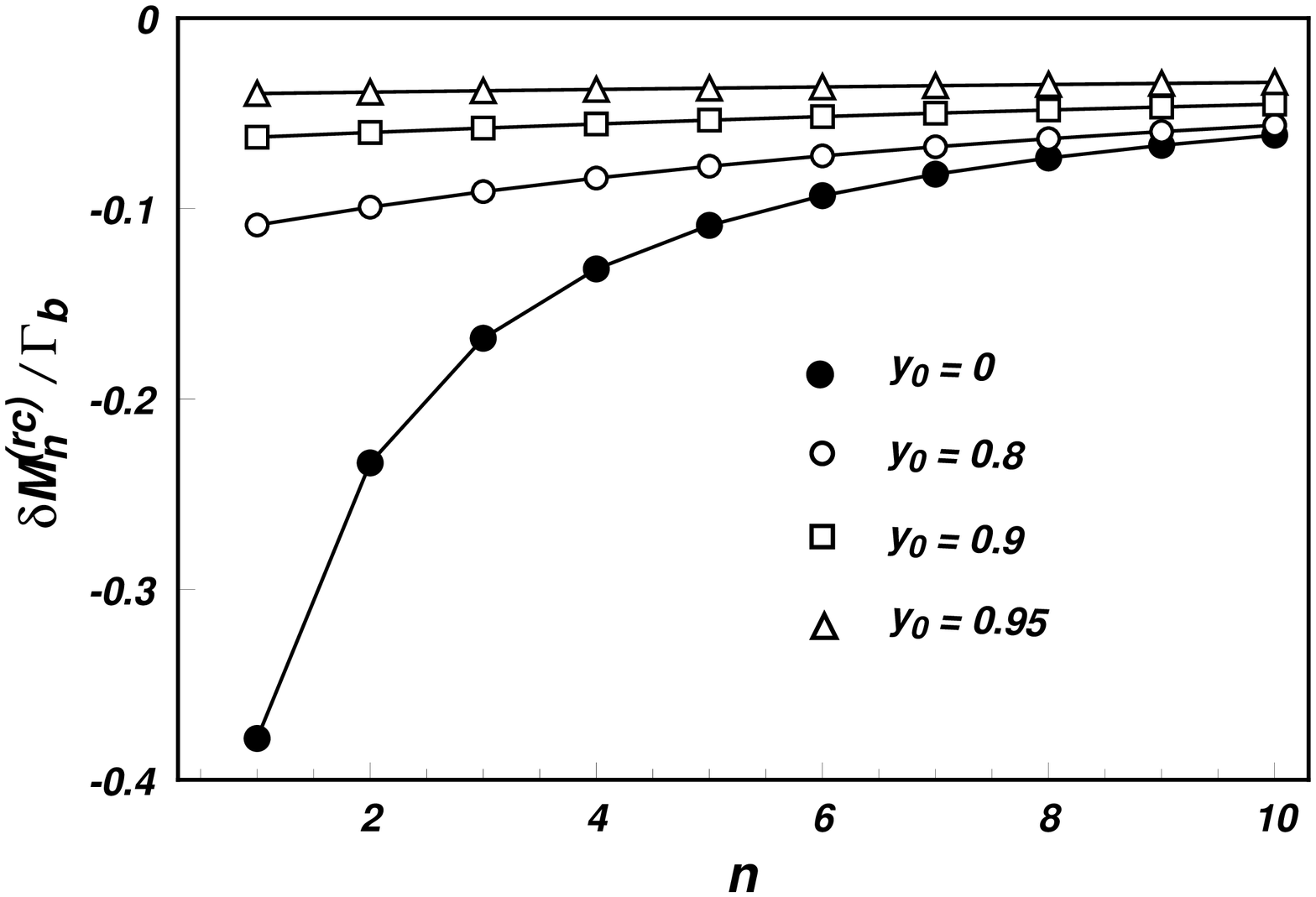}}
   \epsfysize=9cm
   \centerline{\epsffile{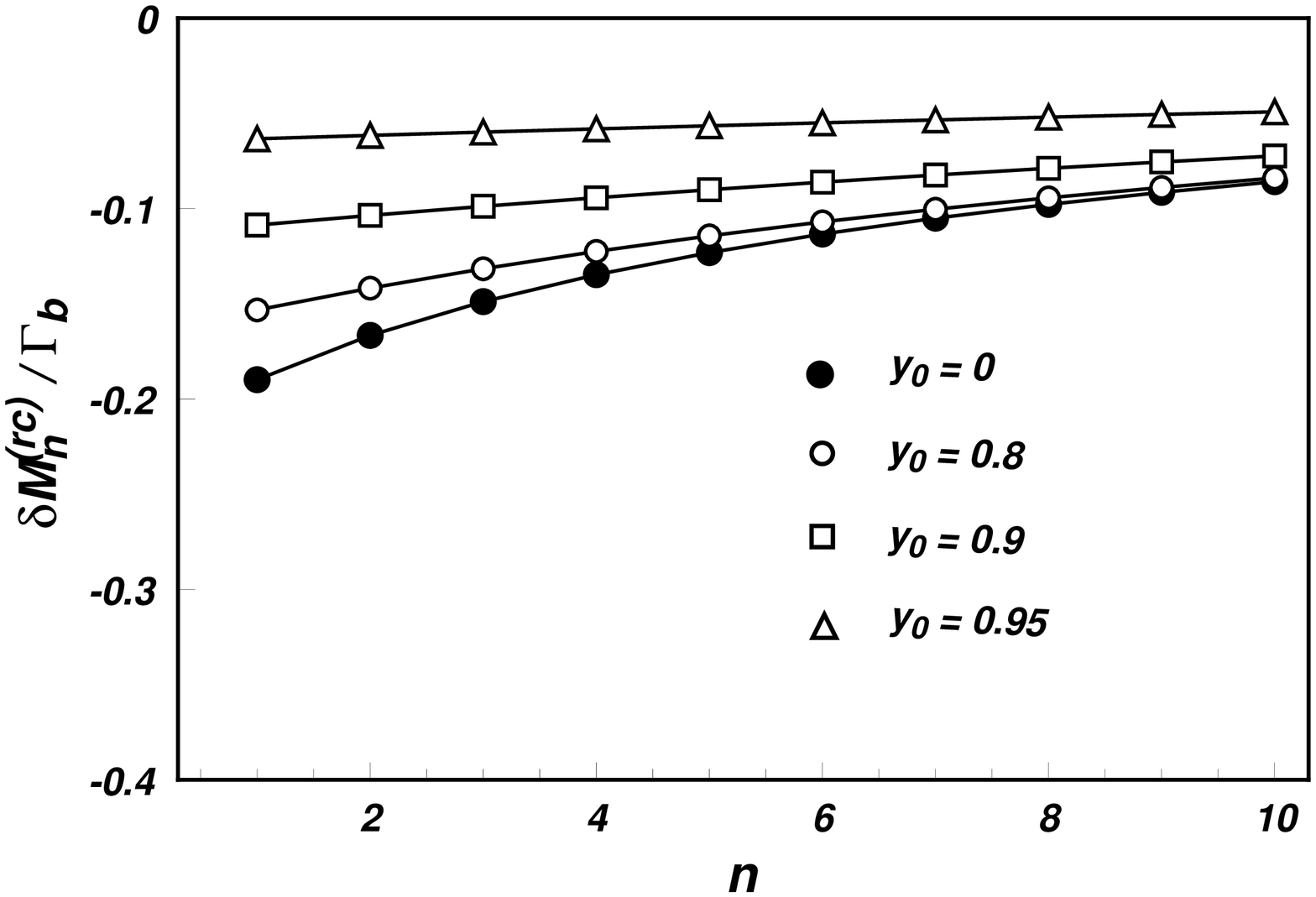}}
   \centerline{\parbox{11cm}{\caption{\label{cutr}
The effect of a lower cut off $y_0$ (c.f.\ (\protect{\ref{mcut}}))
on the radiative corrections
to the moments. The upper figure are the moments obtained from
(\protect{\ref{rc}}) with lower cut off, while the lower plot is
obtained from (\protect{\ref{rc2}}).
   }}}
\end{figure}

The effect of a lower cut in the $y$ integration
for the moments as in (\ref{mcut}) leads
to a doubly logarithmic dependence of the form $\ln^2 (1-y_0)$
on the lower cut $y_0$, and, if $y_0$ comes close to 1, the effects of
the lower cut may become large, forcing us to go to higher $n$.
However, numerically it turns out that with realistic values of
$y_0$ the situation still remains under control. This is shown
quantitatively in fig.\ref{cutr}, where we use expressions (\ref{rc})
(upper figure) and (\ref{rc2}) (lower figure) to estimate the effect
of a lower bound. Again, for the realistic value $y_0 = 0.9$ the
corrections to the moments $4<n<10$ are less than ten per cent.

Our main conclusion is thus that one may use the combined expansion
for the moments below $n =10$, which is given by
\begin{eqnarray} \label{final}
{\cal M}_n &=& \Gamma_b \left[ \frac{2}{n+1}
+ \frac{\lambda_1}{3m_b^2} \left(\frac{10}{n+1}- 1 - n \right)
+ \frac{\lambda_2}{m_b^2} \left(\frac{22}{n+1} -11 \right) \right.
\\
&& - \left.
\frac{4 \alpha_s (m_b) }{3 \pi (n+1)} \left(\frac{5}{4} + \pi^2 \right)
- \frac{4 \alpha_s (m_b) }{3 \pi} \left(\Delta (n) + \frac{31}{6} \sigma(n)
\right) \right]
\nonumber \\
&& + {\cal O} \left(\left(\frac{\bar\Lambda}{m_b}\right)^3, \alpha_s^2 (m_b),
                    \left(\frac{\bar\Lambda}{m_b}\right)^2 \alpha_s (m_b)
                    \right) \nonumber
\end{eqnarray}
where we have defined
\begin{eqnarray}
\Delta (n) &=&
 \sum_{k=0}^n
\left( \begin{array}{c} n \\ k \end{array} \right) (-1)^k
\left( \frac{1}{k+1} \right)^3
 =  \frac{1}{n+1}
\left[ \left( \sum_{k=1}^{n+1} \frac{1}{k} \right)^2 +
       \sum_{k=1}^{n+1}\frac{1}{k^2} \right] \\
&& \longrightarrow \frac{1}{n} \ln ^2 n \mbox{ for } n \gg 1
\nonumber \\
\sigma (n) &=& \sum_{k=0}^n
\left( \begin{array}{c} n \\ k \end{array} \right) (-1)^{k+1}
\left( \frac{1}{k+1} \right)^2 =
- \frac{1}{n+1} \sum_{k=1}^{n+1} \frac{1}{k}
\longrightarrow -\frac{1}{n} \ln n \mbox{ for } n \gg 1
\end{eqnarray}
The values for $\Delta (n)$ and $\sigma (n)$ are tabulated in
tab.\ref{tab1}.

Furthermore, the fact that data will be available only for $y > y_0
\sim 0.9$ only has a small effect on the moments in the range of $n$
considered; the moments including
a lower cut $y_0$ may still be treated in the combined expansion.

\begin{table}
\begin{center}
\begin{tabular}{|c|c|c|}
\hline
$n$    &  $\Delta (n)$  &  $\sigma (n)$  \\
\hline
 0     &    2.000       &   $-$1.000   \\
 1     &    1.750       &   $-$0.750   \\
 2     &    1.574       &   $-$0.611   \\
 3     &    1.441       &   $-$0.521   \\
\hline
 4     &    1.335       &   $-$0.457   \\
 5     &    1.249       &   $-$0.408   \\
 6     &    1.176       &   $-$0.370   \\
 7     &    1.114       &   $-$0.340   \\
 8     &    1.060       &   $-$0.314   \\
 9     &    1.013       &   $-$0.293   \\
 10    &    0.971       &   $-$0.274   \\
\hline
\end{tabular}
\end{center}
\caption{Numerical values for the functions $\Delta (n)$ and
$\sigma (n)$}
\label{tab1}
\end{table}

\begin{figure}[t]
   \vspace{0.5cm}
   \epsfysize=9cm
   \centerline{\epsffile{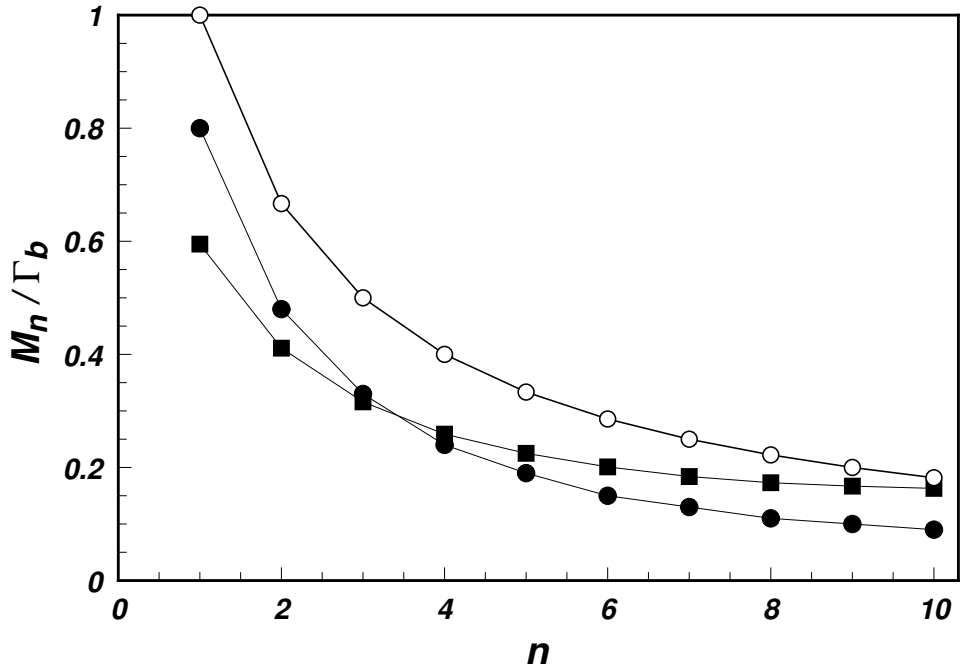}}
   \centerline{\parbox{11cm}{\caption{\label{fig4}
The final result for the moments, including both radiative
and nonperturbative corrections according to (\protect{\ref{final}}).
The solid dots are the result
for $\lambda_1 = -0.3$ GeV${}^2$, the solid boxes are for
$\lambda_1 = -0.6$ GeV${}^2$. For comparison we also plot the
parton model result (open circles).
   }}}
\end{figure}

The final result for the moments is plotted in
Fig.\ref{fig4}, where the moments obtained from the combined
expansion in $\alpha_s (m_b)$ and $\bar\Lambda / m_b$ according to 
(\ref{final}) are shown. The solid dots are the result for ``small''
$\lambda_1 = -0.3$ GeV${}^2$, and the solid boxes are for ``large''
$\lambda_1 = -0.6$ GeV${}^2$.  

To summarize, one may
analyze the moments up to 12 before  the corrections become too big,
of order 100$\%$. We also see that nonperturbative corrections are
very small for small values of $\lambda_1$, and key role is
played by the radiative ones. The reason for
the smallness of nonperturbative corrections is an almost complete
cancellation between the terms proportional
to $\lambda_1$ and $\lambda_2$. For the
case $\lambda_1=-0.6$ GeV$^2$ the cancellation is only partial
and nonperturbative correction are sizable and positive.
This indicates that higher
twist effects may play an important role also in the endpoint region,
and taking into account only the leading twist
contribution, corresponding to the function $F$, is not enough.
This remark, however, applies to a precise description of the
spectrum in the endpoint region, while
for the moments with sufficiently small $n$ it is safe to
use the combined expansion in $\alpha_s (m_b)$ and $\bar\Lambda / m_b$,
which contains pieces of nonleading twist in the constant terms and the
contributions decreasing with $n$.

\section{Conclusions}
Since data on $b \to u$ transitions
will be restricted to a small window between 2.3 and 2.6
GeV, it is mainly the endpoint of the spectrum which will be accessible
in experiment. On the other hand, the endpoint region is the most
difficult region from the theoretical point of view, since here
large radiative corrections are entangled with large nonperturbative
effects. In fact, it is not even obvious, whether and how one may
distinguish the two sources of corrections close to the endpoint.

The main result of this paper is that one may define suitable
averages of the inclusive decay distribution of semileptonic $b \to u$
transitions, which on one hand may be calculated reliably, and which on
the other hand are mainly sensitive in the window between the
kinematic endpoints of $b \to c$ and $b \to u$ semileptonic decays.
We have concentrated on moments of the energy spectrum of the charged
lepton and have performed a detailed numerical analysis, which
shows that moments for $n < 10$ may be systematically
calculated in a combined $\alpha_s (m_b)$ and $\Lambda / m_b$
expansion; both types of corrections may be studied systematically.
Furthermore, the moments with $n > 4$ are sensitive mainly in the
experimentally accessible window. Thus there is indeed a region
$4 < n < 10$ for which the moments defined above may be useful
to perform a determination of $|V_{ub}|$.

In fact, such a combined expansion is even valid for moments defined
with a lower cut off $y_0$, as given in (\ref{mcut}) as long as $y_0$
is not too close to one. Furthermore, for a realistic value $y_0 = 0.9$
the deviations of the moments including a cut from the full ones are
small in the region of $n$ considered. Thus for an experimental analysis
one may as well deal with the moments including a cut.

In order to extract $|V_{ub}|$ from these moments one has to know
the two matrix elements $\lambda_1$ and $\lambda_2$. While $\lambda_2$
is known from the hadron spectrum, $\lambda_1$ is still uncertain;
we have used typical ``large'' and ``small'' values, but one may hope
to extract this parameter from experiment eventually.

One possibility to perform a determination of $\lambda_1$ is in fact
to use the moments defined above. The ratio of two moments
will depend only on $\lambda_1 / m_b^2$, $\lambda_2 / m_b^2$ and
$\alpha_s (m_b)$. The strong coupling and $\lambda_2$ are known,
and one may in principle extract $\lambda_1$ and $m_b$ from two
ratios of moments. After having done this one may then use any one
of the moments to obtain $\Gamma_b$ and hence $|V_{ub}|$. To
proceed along these lines one needs a measurement of three moments
in the range $4 < n < 10$; any measurement of additional moments
may be used to cross check the $n$ dependence of the moments.

The idea presented here may be refined in various aspects, and we
consider this paper as a first try to study whether one may extract
$|V_{ub}|$ using the moments of the decay distributions. As far as the
radiative corrections are concerned, we have only considered the
singular and nonvanishing terms close to the endpoint. In this point
one may refine the analysis by taking into account the complete
corrections, which may be found in the literature. Furthermore, also
the corrections of order $\bar\Lambda / m_b$ have been considered
\cite{Bigiii} and may be also included in the analysis. However,
we do not expect that any of these higher order corrections will change
any of our conclusions.

Finally, one may also think of considering other averages of decay
distributions and generalize the moments according to
\begin{equation}
\widetilde{{\cal M}}_n = \int\limits_0^{1+\bar\Lambda / m_b}  dy  \,\,
{\cal F}_n (y) \frac{d\Gamma}{dy}
\end{equation}
where ${\cal F}_n$ is some set of functions. A good choice of ${\cal F}_n$
may possibly allow a more reliable calculation of the corresponding
moments $\widetilde{{\cal M}}_n$ , and on he other hand be more sensitive
to the experimentally accessible region.

\section*{Acknowledgements}
We want to thank the theory group of SLAC
and especially M. Peskin and S. Brodsky
 for the hospitality
extended to both of us. It was a pleasure to discuss issues concerning
this paper with S. Brodsky,  M. Peskin
 and H. Anlauf. B.B. wants to thank M. Shifman and T.M. wants to
thank D. Cassel, M. Neubert and N. Uraltsev for discussions.

\end{document}